\documentclass[12pt]{article}
\usepackage{graphics,epsfig}
\usepackage[english]{babel}
\newcommand{\cinst}[2]{$^{\mathrm{#1}}$~#2\par}
\newcommand{\crefi}[1]{$^{\mathrm{#1}}$}

\begin{document}


\begingroup

\vglue.5cm


\vspace{1.3cm}
\begin{center}

{\Large{\bf The total yields of $K^{+}(890)$, ${\Sigma}^{+}(1385)$
and ${\Sigma}^0$ in  \\ \vspace{0.5cm} neutrino-induced reactions
at $<E_{\nu}> \approx$ 10~GeV}}
\end{center}

\vspace{1.cm}

\begin{center}
{\large SKAT Collaboration}

 N.M.~Agababyan\crefi{1}, V.V.~Ammosov\crefi{2},
 M.~Atayan\crefi{3},\\
 N.~Grigoryan\crefi{3}, H.~Gulkanyan\crefi{3},
 A.A.~Ivanilov\crefi{2},\\ Zh.~Karamyan\crefi{3},
V.A.~Korotkov\crefi{2}

\setlength{\parskip}{0mm}
\small

\vspace{1.cm} \cinst{1}{Joint Institute for Nuclear Research,
Dubna, Russia} \cinst{2}{Institute for High Energy Physics,
Protvino, Russia} \cinst{3}{Yerevan Physics Institute, Armenia}
\end{center}
\vspace{100mm}

{\centerline{\bf YEREVAN  2005}}

\newpage
\begin{abstract}

Using the data obtained with SKAT bubble chamber, the total yields
of $K^+(892)$, $\Sigma^+(1385)$ and $\Sigma^0$ are estimated for
the first time in neutrino-induced reactions at moderate energies
($<E_{\nu}>$ = 10.4~GeV). It is shown, that the recently observed
\cite{ref1,ref2} enhancement of the $K^0$ and $\Lambda$ yields in
$\nu A$ interactions (as compared to $\nu N$ interactions) is
contributed only slightly by the $K^+(892)$ and $\Sigma^+(1385)$
production, respectively. The decay contribution to the $K^0$ and
$\Lambda$ yields is found to be in qualitative agreement with
higher energy ($<E_\nu> \geq$\, \, 40~GeV) data. It is shown, that
the energy dependence of the $K^+(892)$ mean multiplicity in $\nu
N$ interactions is approximately linear in the range of $<E_{\nu}>
\approx$ 10-60~GeV, while that for $\Sigma^0$ in $\nu A$
interactions (for $A$ = 20-21) is approximately logarithmic in the
range of $<E_{\nu}> \approx$  10-150~GeV.

\end{abstract}
\newpage

As it has been shown recently \cite{ref1,ref2}, the total mean
multiplicity of neutral strange particles ($K^0$, $\Lambda$) in
neutrinonuclear interactions exceeds noticeably that in ${\nu}N$
interactions. As a result, the ratio of yields of strange and
non-strange mesons, $R(K^0/{\pi}^-)$, exhibits a significant A-
dependence \cite{ref2}. While the multiplicity gain for $\Lambda$
is shown to be dominated by secondary intranuclear processes
${\pi}N \rightarrow {\Lambda}X$, only a small fraction of that for
$K^0$ can be attributed to secondary reactions ${\pi}N \rightarrow
K^0 X$ \cite{ref2}. In this context, it is interesting to
investigate the nuclear medium influence on the total yields of
other strange particles, in particular, of ${\Sigma}^0$ hyperon
and strange resonances ${\Sigma}(1385)$ and $K(892)$, in
neutrino-induced reactions. \\ Hitherto, the total yield of
${\Sigma}^0$ is measured only in ${\nu}A$ interactions at $A$ = 20
\cite{ref3,ref4,ref5} and $A \approx$ 12 \cite{ref6}, while those
for $\Sigma(1385)$ and $K(892)$ are extracted for ${\nu}N$
\cite{ref7,ref8,ref9}, as well as for ${\nu}A$ interactions at $A
\approx$ 12 \cite{ref6}. These studies are performed at relatively
high energies, the mean neutrino energy being in the range of
$<E_{\nu}>$ = 40-150 GeV. \\ In this work, the first attempt is
undertaken to estimate the total yields of ${\Sigma}^0$,
${\Sigma}^+(1383)$ and $K^+(892)$ in neutrino-induced reactions at
moderate energies, $<E_{\nu}> \approx$ 10~GeV. To this end, the
data from the SKAT bubble chamber \cite{ref10}, filled with a
propane-freon mixture, were used (see \cite{ref2} and references
therein). The total number of accepted events with the invariant
mass of the hadronic system $W >$ 1.8~GeV (exceeding the threshold
value for the strange resonance production) was equal to 4888, the
corresponding mean neutrino energy being equal to $<E_{\nu}>$
10.4~GeV. The selection criteria for the decay of neutral strange
particles and the procedure of their identification were similar
to those applied in \cite{ref11}. The number of accepted neutral
strange particles ($V^0$'s) was 104 out of which 44(60) had the
biggest probability to be identified as $K^0(\Lambda)$. The
corresponding average multiplicities, corrected for the decay
losses, are $<n_{V^0}> = (8.13 \pm 0.80)\cdot 10^{-2}$, $<n_{K^0}>
= (4.97 \pm 0.75)\cdot 10^{-2}$ and $<n_{\Lambda}> = (3.21 \pm
0.41)\cdot 10^{-2}$. \\ For the further analysis the whole event
sample was subdivided, using several topological and kinematical
criteria, in the 'quasinucleon' subsample ($A$ = 1) and the
nuclear subsample, the latter having an effective atomic weight $A
\approx$ 21 (see \cite{ref2,ref12} for details). The mean
multiplicities of $K^0$ and $\Lambda$ in these subsamples are,
respectively, $<n_{K^0}>_N = (3.44 \pm 0.86)\cdot 10^{-2}$, \,
$<n_{K^0}>_A = (5.07 \pm 0.76)\cdot 10^{-2}$ and $<n_{\Lambda}>_N
= (2.19 \pm 0.50)\cdot 10^{-2}$, $<n_{\Lambda}>_A = (3.33 \pm
0.43)\cdot 10^{-2}$. \\ The effective mass distributions for
systems ${\Lambda}\gamma$, $\Lambda{\pi^+}$ and $K^0{\pi^+}$,
plotted in Fig. 1, indicate on signals (being, do to the
restricted statistics, of low confidence level) near the masses
of, respectively, $\Sigma^0$, $\Sigma^+(1385)$ and $K^+(892)$ for
the nuclear subsample and of $K^+(892)$ for the 'quasinucleon'
subsample. No signals are seen for the systems ${\Lambda}\gamma$
and $\Lambda{\pi^+}$ in the 'quasinucleon' subsample (not shown).
The dashed curves in Fig. 1 represent the background distributions
obtained by particle combinations from different events of similar
topologies and normalized to the experimental distributions
outside the signal region. The distribution on the
$\Lambda{\gamma}$ effective mass is corrected for the efficiency
of the $\gamma$ detection. The effective mass resolution for the
systems ${\Lambda}\gamma$, $\Lambda{\pi^+}$ and $K^0{\pi^+}$
around the the signal region is equal to, respectively, 40, 58 and
47 MeV. The distributions were fitted as a sum of the fixed
background distribution and a Gaussian one (for the case of
$\Lambda{\gamma}$) or a relativistic Breit-Wigner distribution
(for the case of $\Lambda{\pi^+}$ and $K^0{\pi^+}$) taking into
account the experimental mass resolution (see the solid curves in
Fig. 1). \\ The extracted mean multiplicities of $K^+(892)$,
$\Sigma^+(1385)$ and $\Sigma^0$, as well as those for $K^0$ and
$\Lambda$, are presented in Table 1. The upper limits of
$<n(\Sigma^+(1385)\rightarrow \Lambda{\pi^+)>_N}$ and
$<n(\Sigma^0)>_N$ for 'quasinucleon' interactions are estimated by
simply subtraction of the normalized background distribution from
the experimental one. The last column of Table 1 shows the nuclear
multiplicity gain, $\delta_h = <n_h>_A - <n_h>_N$.

\begin{table}[ht]
\begin{center}
\begin{tabular}{|l|c|c|c|}
  \hline
&&&\\ { Particle}&{$<n>_{N}$}&{$<n>_{A}$}&{$<n>_{A} - <n>_{N}$}
\\ \hline  &&& \\ {$K^0$}&{
3.44$\pm$0.86}&{5.07$\pm$0.76}& {1.63$\pm$0.86}
\\ &&& \\ {$K^{+}(892) \rightarrow K^0 {\pi^+}$}&{
0.88$\pm$0.53}&1.02$\pm$0.48&0.14$\pm$0.53
\\ &&&  \\ \hline &&& \\ $\Lambda$&
2.19$\pm$0.50&3.33$\pm$0.43&1.14$\pm$0.48\\ &&&
\\ { ${\Sigma}^{+}(1385) \rightarrow \Lambda
{\pi^+}$}&$< $0.13 &0.39$\pm$0.21& $<$0.39$\pm$0.21
\\ &&& \\ \hline &&& \\{ ${\Sigma}^0$}& $<$
0.53&0.82$\pm$0.45 &$<$ 0.82 $\pm$ 0.45 \\ &&&
\\ \hline
\end{tabular}
\end{center}
\caption{The mean multiplicities $<n>_N$ and $<n>_A$ of strange
particles and the multiplicity gain $<n>_A - <n>_N$ (in
10$^{-2}$).}
\end{table}

\noindent For the first time, an indication is obtained that this
gain for resonances, $\delta_{K^+(892)}$ and
$\delta_{\Sigma^+(1385)}$, plays only a minor role in that for
daugther particles, being significantly smaller than
$\delta_{K^0}$ and $\delta_{\Lambda}$, respectively.
\\ The decay contribution to the mean multiplicities of $K^0$ and
$\Lambda$ is quoted in Table 2

\begin{table}[h]
\begin{center}
\begin{tabular}{|l|c|c|}
  \hline && \\
 Ratio&${\nu}$N& ${\nu}A$
 \\ \hline  &&  \\  $K^{+}(892) / K^0$&
0.26$\pm$0.17&0.20$\pm$0.10   \\ && \\ \hline &&
 \\  ${\Sigma}^{+}(1385)
/ {\Lambda}$ & $< $0.08 &0.11$\pm$0.06 \\ &&
\\ \hline  && \\ ${\Sigma}^0 / {\Lambda}$ &  $<$
0.18 & 0.24$\pm$0.14 \\ &&
\\ \hline

\end{tabular}

\end{center}
\caption{The decay contribution to the mean multiplicities of
$K^0$ and $\Lambda$.}
\end{table}

\noindent and compared with the available data
\cite{ref3,ref4,ref5,ref6,ref7,ref8,ref9} in Fig. 2. The data of
Fig. 2 do not indicate on a noticeably energy or $A$- dependences
for the ratios $R(K^+(892)/K^0)$ and $R(\Sigma^0/\Lambda)$ which
are more or less consistent with averaged values 0.16$\pm$0.01 and
0.09$\pm$0.02, respectively (see dashed lines in Figs. 2a and 2c),
while the data on $R(\Sigma^+(1385)/\Lambda)$ demonstrate larger
discrepancy, mainly due to the value at $<E_{\nu}>$ = 45~GeV for
$A \approx$ 12 \cite{ref6}. \\ The energy dependence of
$<n(K^+(892))>_N$ (corrected for the decay mode $K^+(892)
\rightarrow K^+{\pi^0}$) and $<n(\Sigma^0)>_A$ (at $A \approx$
20-21) is plotted in Fig. 3. As it is seen, the data on
$<n(K^+(892))>_N$ in the energy range $<E_{\nu}>\approx$ =
10$-$60~GeV can be described by a linear dependence
$<n(K^+(892))>_N = b\cdot (<E_\nu> - E_\nu^{thr}$) with $b =
(1.12\pm0.12)\cdot 10^{-3}/$GeV, while the energy rise of
$<n(\Sigma^0)>_A$, being much weaker, is consistent, in the range
of $<E_{\nu}>\approx$ = 10$-$150~GeV, with a logarithmic
dependence, $<n(\Sigma^0)>_A = a \cdot \ln(<E_\nu>/E_\nu^{thr})$
with $a = (2.60\pm0.56) \cdot 10^{-3}$, where $E_\nu^{thr}$ is the
threshold neutrino energy for the corresponding resonance
production. \\

\noindent In conclusion, the total yields of $K^+(892)$,
$\Sigma^+(1385)$ and $\Sigma^0$ are estimated for the first time
in neutrino-induced reactions at moderate energies ($<E_\nu>
\approx$ 10~GeV). The contribution from the decay of those
particles to the yields of $K^0$ and $\Lambda$ is found to be
rather small, being in a qualitative agreement with other, higher
energy ($<E_\nu> \geq$ 40~GeV) data. It is shown, that only a
small fraction of the nuclear enhancement of the $K^0$ and
$\Lambda$ yields (observed recently in \cite{ref1,ref2}),
$<n_K^0>_A - <n_K^0>_N$ and $<n_{\Lambda}>_A - <n_{\Lambda}>_N$,
respectively, can be contributed by the $K^+(892)$ and
$\Sigma^+(1385)$ production. It is shown, that in a wide energy
range, the energy dependence of $<n(K^+(892))>_N$ in ${\nu}N$
interactions is approximately linear, while that of
$<n(\Sigma^0)>_A$ in ${\nu}A$ interactions (for $A \approx$ 20-21)
is consistent with the logarithmic one.

\noindent {\bf{Acknowledgement:}} The activity of one of the
authors (H.G.) is partly supported by Cooperation Agreement
between DESY and YerPhI signed on December 6, 2002. The autors
from YerPhI acknowledge the supporting grants of Calouste
Gulbenkian Foundation and Swiss Fonds "Kidagan".


\newpage
\begin{figure}
\resizebox{1.1\textwidth}{!}{\includegraphics*[bb=50 120 600
530]{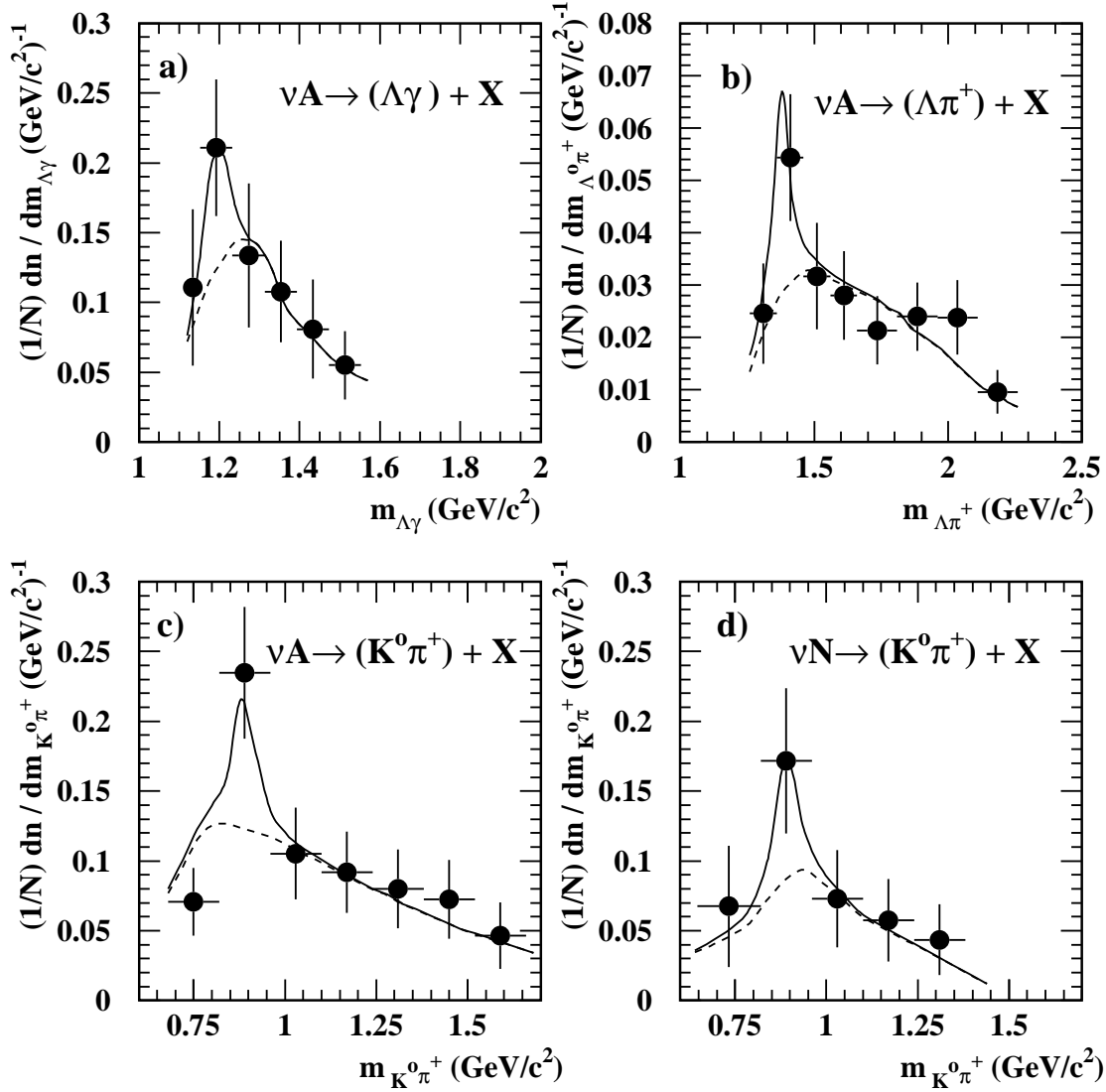}}  \vspace{1.5cm} \caption{The effective mass
distributions for systems ${\Lambda}\gamma$, $\Lambda{\pi^+}$ and
$K^0{\pi^+}$.}
\end{figure}

\newpage

\begin{figure}
\resizebox{1.1\textwidth}{!}{\includegraphics*[bb=50 120 600
530]{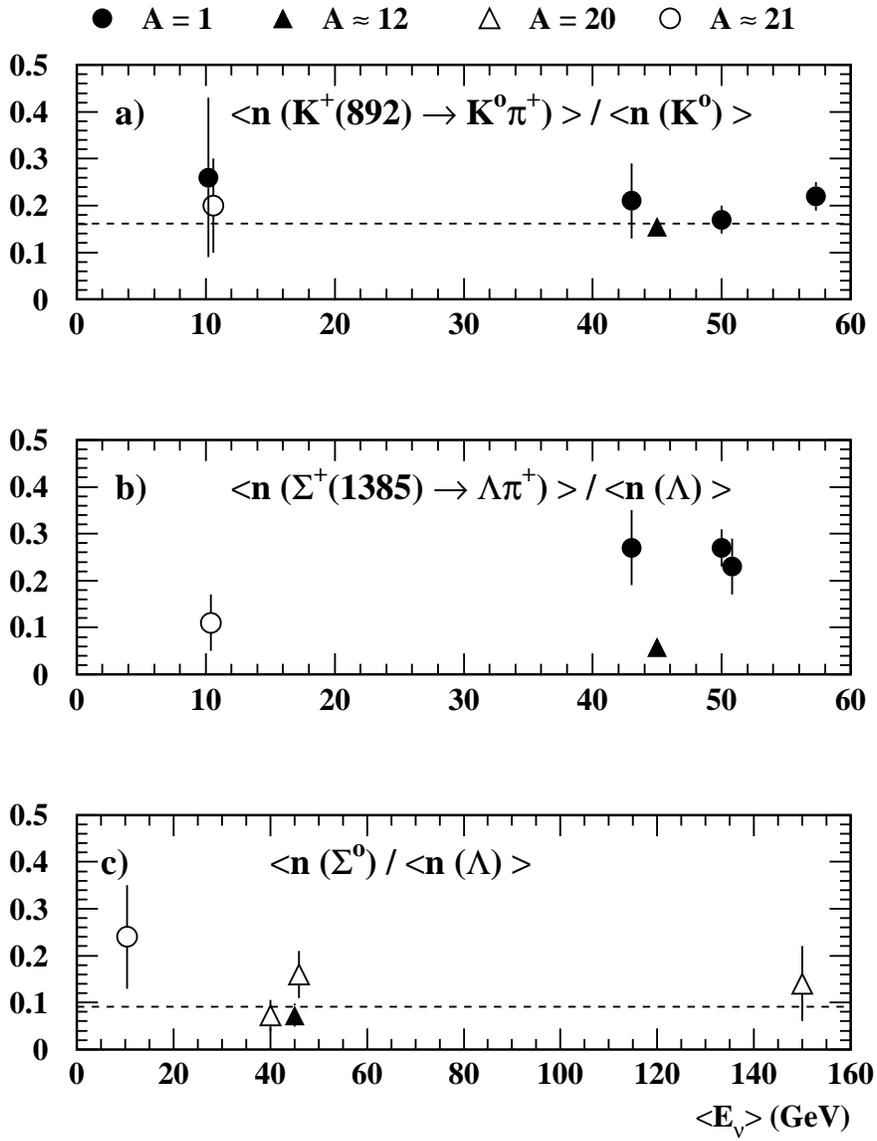}}  \vspace{2cm} \caption{The $E_\nu$- dependence of
the decay contribution to the total yields of $K^0$ and
$\Lambda$.}
\end{figure}

\newpage

\begin{figure}
\resizebox{1.1\textwidth}{!}{\includegraphics*[bb=50 120 600
530]{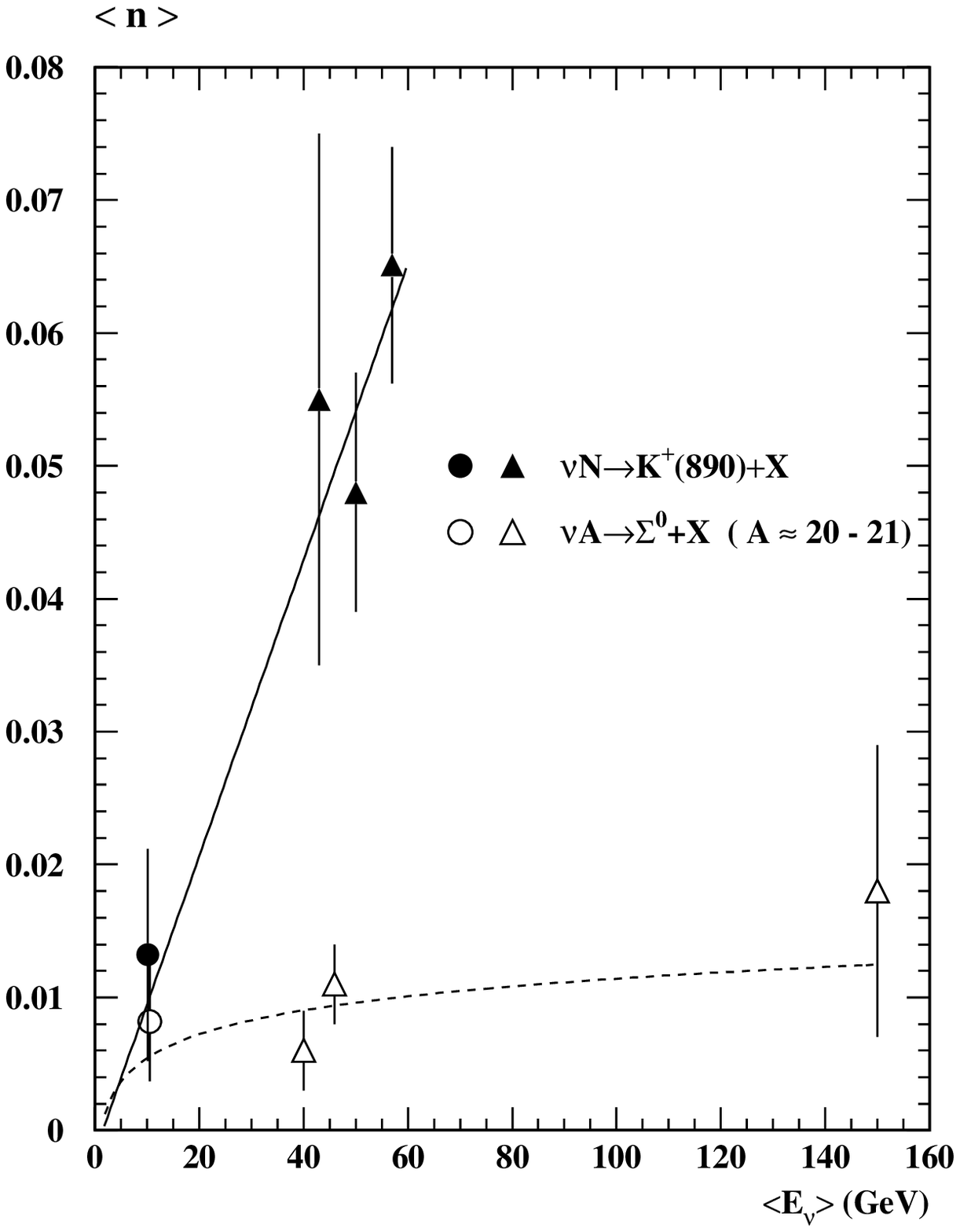}}  \vspace{1.5cm} \caption{The $E_\nu$- dependence
of $<n(K^+(892))>_N$ and $<n(\Sigma^0)>_A$ for $A \approx$ 20-21.}
\end{figure}


\begin{thebibliography}{00}
\parskip=0.pt \parsep=0.pt \itemsep=0.pt

\bibitem{ref1}
N.M.Agababyan et al., (SKAT Coll.), YerPhI Preprint-1593(3), 2004,
Yerevan; Preprint hep-ex/0405076, YAF [Phys. of At. Nucl.] (in
print)
\bibitem{ref2}
N.M.Agababyan et al., (SKAT Coll.), YerPhI Preprint- ?, 2005,
Yerevan; Preprint hep-ex/?
\bibitem{ref3}
N.J.Baker et al., Phys. Rev. D {\bf34}, 1251, 1986
\bibitem{ref4}
S.Willocq et al. (WA59 Coll.), Z. Phys. C {\bf53}, 207, 1992
\bibitem{ref5}
D.DeProspo et al., (E623 Coll.), Phys. Rev. D {\bf50}, 6691, 1994
\bibitem{ref6}
P.Astier et al., (NOMAD Coll.), Nucl. Phys. B {\bf621}, 3, 2002;
hep-ex/0111057
\bibitem{ref7}
H.Gr$\ddot{a}$ssler et al., (BEBC Coll.), Nucl. Phys. B {\bf194},
1, 1982
\bibitem{ref8}
D.Allasia et al., (BEBC Coll.), Nucl. Phys. B {\bf224}, 1, 1983
\bibitem{ref9}
G.T.Jones et al., (WA21 Coll.), Z. Phys. C {\bf57}, 197, 1993
\bibitem{ref10}
V.V.Ammosov et al., Fiz. Elem. Chastits At. Yadra {\bf23}, 648,
1992 [Sov. J. Part. Nucl. {\bf23}, 283, 1992]
\bibitem{ref11}
V.V.Ammosov et al., (SKAT Coll.), Z. Phys. C {\bf30}, 183, 1986
\bibitem{ref12}
N.M.Agababyan et al., (SKAT Coll.), YerPhI Preprint-1578(3), 2002,
Yerevan

\end{thebibliography}
\end{document}